\documentclass[%
 aip,
 amsmath,amssymb,
 reprint,%
]{revtex4-1}
\usepackage{graphicx}
\usepackage{dcolumn}
\usepackage{bm}
\usepackage[latin1]{inputenc}
\usepackage[T1]{fontenc}
\usepackage{mathptmx}
\usepackage{etoolbox}
\usepackage{subfigure}

\makeatletter
\def\@email#1#2{%
 \endgroup
 \patchcmd{\titleblock@produce}
  {\frontmatter@RRAPformat}
  {\frontmatter@RRAPformat{\produce@RRAP{*#1\href{mailto:#2}{#2}}}\frontmatter@RRAPformat}
  {}{}
}%
\makeatother
\begin{document}

\preprint{AIP/123-QED}

\title{Simulation of exceptional-point systems on quantum computers for quantum sensing}
\author{Chetan Waghela}
\email{chetan.waghela@iitrpr.ac.in}
\author{Shubhrangshu Dasgupta}%
\affiliation{ 
Department of Physics, Indian  Institute of Technology Ropar, Rupnagar, Punjab-140001, India.
}%


\date{\today}

\begin{abstract}
There has been debate around applicability of exceptional points (EP) for quantum sensing. To resolve this, we first explore how to experimentally implement the nonhermitian non-diagonalizable Hamiltonians, that exhibit EPs, in quantum computers which run on unitary gates. We propose to use an ancilla-based method in this regard. Next, we show how such Hamiltonians can be used for parameter estimation using quantum computers and analyze its performance in terms of the Quantum Fisher Information ($QFI$) at EPs, both without noise and in presence of noise. It is well known that $QFI$ of a parameter to be estimated is inversely related to the variance of the parameter by the quantum Cramer-Rao bound. Therefore the divergence of the $QFI$ at EPs promise sensing advantages. We experimentally demonstrate in a cloud quantum architecture and theoretically show, using Puiseux series, that the $QFI$ indeed diverges in such EP systems which were earlier considered to be non-divergent. 
\end{abstract}

\maketitle

%

\section{\label{sec:1}Introduction}

Quantum operations can be classified as either unitary or non-unitary. Much attention has been given to unitary operations in physics. In fact, in a quantum computer, quantum gates are made up of unitary operations.  However, there are several reasons to consider non-unitary transformations. Open quantum systems exhibit non-unitary dynamics \cite{PhysRevA.83.062317,wei2016duality,zheng2021universal,lloyd1996universal,zheng2021universal}. In quantum chemistry, as well, non-unitary evolutions are often studied \cite{baskaran2022adapting,Jnane:2020swk,benfenati2021improved}. Quantum speedups have also been proposed in nonlinear quantum computing through non-unitary Abrams-Lloyds gate \cite{abrams, nonlinear}. Quantum machine learning needs non-unitarity and non-linearity \cite{cong, beer, gili, schuld} too.

Consider a transformation, represented  by $e^{-iA}$. If $A$ is hermitian, this transformation  becomes unitary and vice-versa. However, for  $e^{-iA}$ to be non-unitary, its generator $A$ must be non-hermitian. 
This requires one to study the behaviour non-hermitian matrices. Usually, non-hermiticity does not guarantee that its eigenvalues would be real, unlike the hermitian ones.  In 1998, \cite{bender1998real} showcased that there exist certain non-hermitian Hamiltonians that have real eigenvalues. 

This sparked a debate around postulates of quantum mechanics and explorations of such systems in quantum domain. It was shown in \cite{mostafazadeh2010pseudo} that all such Hamiltonians, if diagonalizable, are related to hermitian ones by similarity transformations, and hence are ``pseudo-hermitian". On the other hand, the non-hermitian Hamiltonians, which are non-diagonalizable,  lie at degeneracy called the non-hermitian degeneracy or exceptional points (EP) in literature. Such an operator has degeneracy in both eigenvalues and eigenvectors of the operator. In this paper, we will consider these non-diagonalizable 
systems in the context of quantum sensing, particularly for estimation of an unknown parameter.

For the best possible estimate of any parameter $\gamma$, there exists 
the so-called Quantum Cramer-Rao Bound (QCRB)  \cite{braunstein1994statistical,paris2009quantum} which is given by the inverse of the quantum Fisher information ($QFI$). This can be represented as 
\begin{equation}
    \Delta \gamma \leq \frac{1}{\sqrt{QFI_{\gamma}}}\;.
    \label{delgam}
\end{equation}
It can be observed that higher the $QFI$, less the variance ($\Delta \gamma^2$) and hence more precise the estimation. Hence, a very high $QFI$ shows promise in its exploitation for precise sensing. 

It was first theoretically pointed out in \cite{brody2013information} that the EPs can be utilized for parameter estimation as $QFI$ may diverge at EP of a non-hermitian non-diagonalizable Hamiltonian, in absence of noise. Note that the divergence of $QFI$ does not violate Heisenberg's uncertainty principle as the latter is concerned with simultaneous measurement of two canonically conjugate physical variables while QCRB is concerned with precision of measurement of a particular physical variable \cite{toth2022uncertainty, luo2000quantum}.
Later, it was deduced in  \cite{chen2019sensitivity} that $QFI$ does not necessarily exhibit any divergence at EP, and this result was negated by a more accurate analysis in  \cite{zhang2019quantum} that shows that the $QFI$ indeed diverges. This debate of divergence is also highlighted in \cite{duggan2022limitations, wiersig2020review}. In fact, the divergence of $QFI$ is related to the expansion of a perturbed eigenfunctions and eigenvalues in Puiseux series rather than Taylor series for any perturbations at EPs, as was already shown in \cite{kato2013perturbation} and which was often overlooked in literature. In this paper, we specifically employ the Puiseux series in calculating the QFI and show that this unambiguously leads to the divergence, as discussed above, at the EP.

The possibility of divergence of the $QFI$ at EP motivates us to explore how to implement non-hermitian dynamics in real quantum computers.  Till recent times, the non-hermitian systems have been mostly explored in optical systems through an analogous connection between Helmholtz equation and Schrodinger equation \cite{miri2019exceptional}. There have been a few experimental considerations of EPs in certain quantum systems too \cite{naghiloo2019quantum, ding2021experimental}. Here, we democratize access to such exotic systems using quantum computers. Note that these computers, which are accessible via cloud, operate unitarily. In this paper, we show that it is indeed possible to implement the nonunitary dynamics, generated by non-hermitian Hamiltonian, in the existing architecture itself. We propose use of  ancillary qubits in this regard. We employ post-selection strategy to extract the state of the EP system, that will be used to calculate the $QFI$ and to evaluate the performance of the system for parameter estimation. We also provide a circuit model that we use in our cloud experiments in the IBM Q Experience platform.

The achievable precision of a quantum parameter estimation protocol is limited by the noise in the system, which could degrade the $QFI$. The early EP sensing protocols did not consider the effect of noise. Later, various studies related to impact of noise on such sensors were reported \cite{Langbein-2018,lau2018fundamental,chen2019sensitivity,zhang2019quantum}. In \cite{lau2018fundamental}, the authors showed that a non-hermitian non-reciprocal system delivers an unbounded signal-to-noise ratio (SNR) and thereby a  better sensing performance compared to the reciprocal ones, even when it operates away from the EP. On the contrary, in this paper, we focus on studying the behavior of the $QFI$, and not the SNR, in presence of noise. Note that, though for certain parameter domain, the $QFI$ becomes numerically identical to the SNR, they are fundamentally different quantities. While the $QFI$ is related to the amount of information that can be extracted from a quantum system about a particular parameter of interest, the SNR is a measure of the relative strength of a signal to the level of background noise. Thus, in the context of quantum sensing, the $QFI$ is a more suitable quantity than the SNR to characterize the performance of the quantum sensor, more so when the quantum noise dominates over the classical noise and when their sensitivity to changes in a parameter is governed by the quantum rules, e.g., the Heisenberg uncertainty principle.

The structure of the paper is as follows. In the Sec. II, we will first introduce the basic theory of exceptional points in finite dimensional systems. Next, we will describe the  theory related to simulation of non-unitary dynamics using ancilla based method \cite{terashima2005nonunitary}. In the Sec. III, we will present the results of our simulation of dynamics of non-hermitian systems using IBM Q Experience. We will explicitly show the divergence of $QFI$ at exceptional points. Then in Sec. IV,  we discuss effects of various noise models on EP sensors. In Sec. V an example physical quantum sensing system at EP is discussed. We will discuss in Sec.  VI the limitations of the ancilla-based method used in this paper. In Sec. VII, we will summarize our work. We also include three Appendices. In the Appendix A, we describe Puiseux series, which needs to be used to expand the perturbed states of the system at the EP. In the Appendix B, we add a comprehensive discussion of orthogonality, bi-orthogonality and self-orthogonality of the eigenvectors, that are relevant for non-hermitian matrices. In the Appendix C, we derive the explicit  expression of the $QFI$ for non-hermitian systems, that are used in this paper.  


\section{\label{sec:2} Non-hermitian sensing and simulating non-unitary dynamics}

In this Section, we will first provide a brief review of experimental realizations of exceptional points, as will be
useful later. Next, we will describe how one can indeed simulate non-unitary dynamics in unitary-based quantum
computers.

\begin{figure*}
     \centering
     \begin{subfigure}
         \centering
         \includegraphics[width=\textwidth]{ 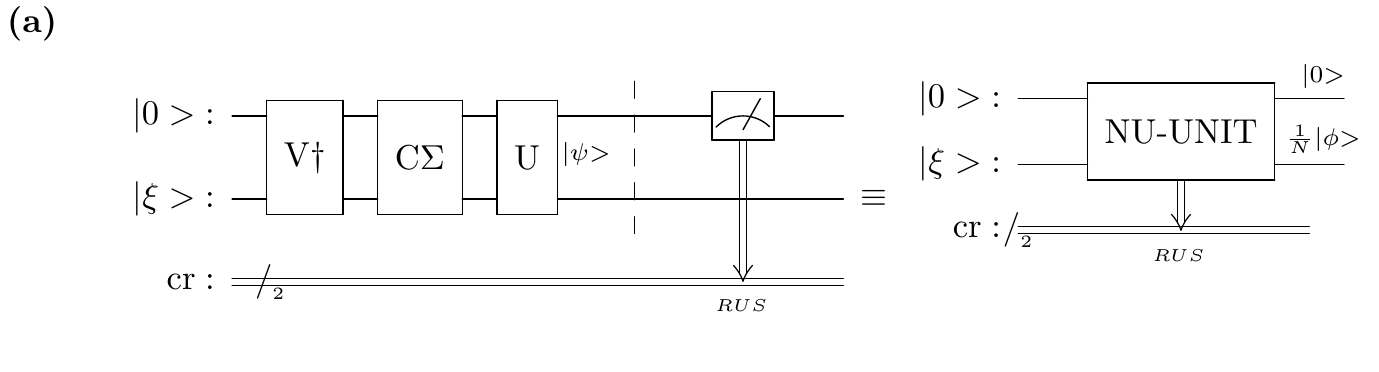}
     \end{subfigure}
     \vfill
     \begin{subfigure}
         \centering
         \includegraphics[width=\textwidth]{ 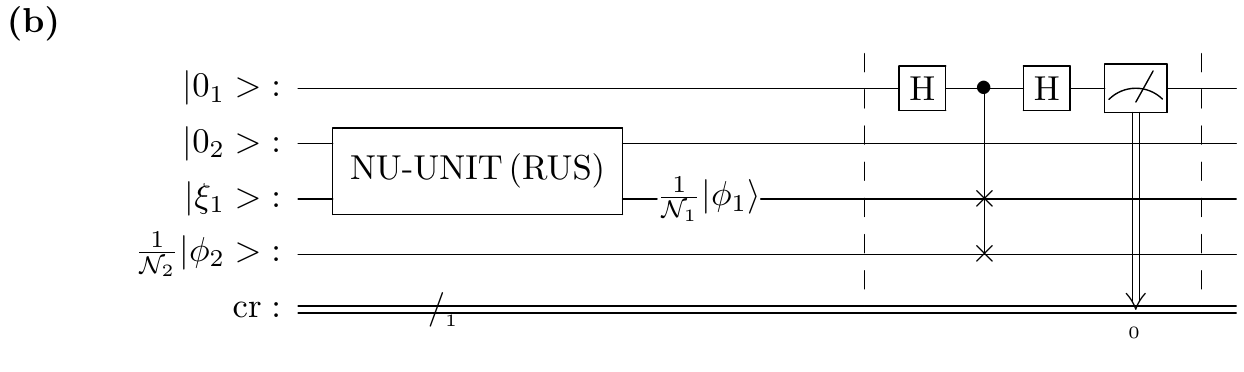}
     \end{subfigure}
     \vfill
\caption{ (a) The circuit diagram for simulating non-unitary evolution. We decompose the non-unitary transformation using SVD transform into $U\Sigma V^{\dagger}$. Then using the procedure as mentioned in the text, we apply the operations on the input state $|\xi,0 \rangle$ to obtain $|\psi \rangle$. This is followed by a post-selection protocol applied on the output using "Repeat until success" (RUS) method. Here we discard all measurements results  which project the ancilla  in the state $|1\rangle$, and  repeat the procedure till we  get the ancilla to be in the state $|0 \rangle$. This collapses the state to $\frac{1}{\cal N}|\phi, 0 \rangle$ which contains the required state due to non-unitary evolution. On the right hand side, we schematically represent the non-unitary gate. Note that 'cr' in the circuit refers to classical register.
(b) The circuit diagram for the SWAP test to measure fidelity between two states i.e. $F=\frac{1}{|\mathcal{N}_1\mathcal{N}_2|^2}|\langle \phi_2|\phi_1 \rangle|^2$. We can consider $\frac{1}{\mathcal{N}_1}|\phi_1 \rangle$ and $\frac{1}{\mathcal{N}_2}|\phi_2 \rangle$ to be the states as given in Eq. \ref{Fidd}. The circuit between the dashed vertical barriers represents SWAP test. We specifically mention $\mathcal{N}$ in denominator to show that non-unitary circuits un-normalize input states and we need to consider normalization of such output states. }
\label{circuit}
\end{figure*}

\subsection{\label{sec:2a} Reviewing experimental demonstrations of exceptional points}

Consider a finite-dimensional system with the Hamiltonian $H$, that can be written as a matrix. It is well known that hermiticity of a matrix guarantees that it has real eigenvalues and it is always diagonalizable. Interestingly, even the non-hermitian matrices, if  diagonalizable and if all of its eigenvalues are real, have been proved in pseudo-hermitian quantum theory to be similar to hermitian matrices. However, this proof does not include the non-diagonalizable (non-hermitian) matrices. A  (non-hermitian) matrix is non-digonalizable iff the algebraic multiplicity of at least one of its eigenvalues is greater than the corresponding geometric multiplicity. This happens only when the system has degeneracy in its eigenvalues as well as eigenvectors, i.e., only when the eigenvalues and eigenvectors coalesce.  Below, we will consider a class of Hamiltonians, as functions of a parameter $\gamma$, that exhibit such coalescence for a certain value of $\gamma$, referring to the EP.  

As mentioned before, there have been several experimental demonstrations of exceptional points in optical systems \cite{miri2019exceptional}. However in the context of quantum system,  their observations have remained rare and elusive \cite{wiersig2020review}. Several applications of exceptional points other than sensing are reported in quantum regime \cite{miri2019exceptional,ozdemir2019parity, zhang2019quantum}. Only a few experimental demonstrations \cite{naghiloo2019quantum,chen2022decoherence} exist where a quantum system, e.g., a superconducting system is used to realize exceptional points. In this paper, we explicitly demonstrate how the exceptional points can be utilized for the purpose of sensing in real quantum computers. We access these exceptional points on a quantum system using the circuit model developed by \cite{terashima2005nonunitary}.

\subsection{\label{sec:2b} Simulating non-unitary dynamics}

Let us consider a finite-dimensional non-hermitian Hamiltonian $NH$ such that $(NH)^{\dagger} \neq NH$. Such Hamiltonians effectively arise in many situations as mentioned in Sec. I. We are largely concerned with a special class of these Hamiltonians, namely the non-diagonalizable ones, as mentioned above. 

\begin{figure*}
    \centering
    \includegraphics[scale=0.8]{ 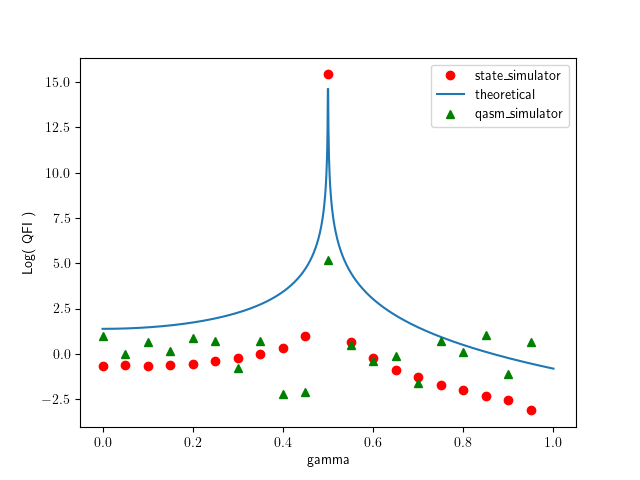}
    \caption{The variation of $\log(QFI_\gamma)$ with respect to  $\gamma$ for the Hamiltonian $NH_1$ (Eq. \ref{Ham}). The $QFI_\gamma$ diverges at the EP, which appears at $\gamma=0.5$. Blue curve represents theoretical plot and red dots indicate simulation values at various $\gamma$ values.}
    \label{result1}
\end{figure*}

\begin{figure*}[ht]
    \centering
    \includegraphics[scale=0.8]{ 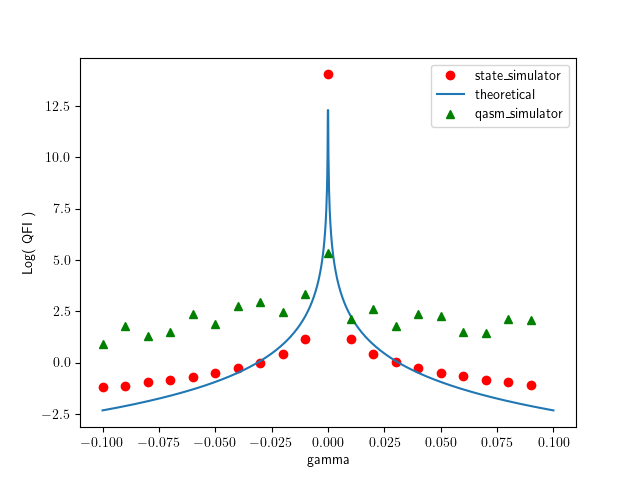}
    \caption{The variation of $\log(QFI)$ with respect to  $\gamma$ for the Hamiltonian $NH_2$ (Eq. \ref{Ham2}). The $QFI$ diverges at the EP, which appears at $\gamma=0$ when $g=|\epsilon|/2$. Blue curve represents theoretical plot and red dots indicate simulation values at various $\gamma$ values.}
    \label{result2}
\end{figure*}

The goal of this simulation is to find the output state $\frac{1}{{\mathcal N}}|\phi\rangle = \exp[\frac{-i (NH) t}{\hbar}]|\xi\rangle $, where $|\xi\rangle$ is an input state and ${\cal N}$ is the normalization constant. Note that, in this notation, the input and output states are both pure states even though the evolution is non-unitary.

We will consider the following steps in this simulation:

(a) {\it Singular value decomposition:} First, we will find the non-unitary time evolution matrix ($NUTE$), where $NUTE=\exp[\frac{-i(NH)t}{\hbar}]$. For simulation purposes, now onward, we will consider $t,\hbar$ = 1. Such a time evolution can be written as a sequence of three separate evolutions using the singular value decomposition (SVD), i.e., $NUTE=U \Sigma V^{\dagger}$, 
where $U$ and $V^{\dagger}$ are always unitary and hence can be simulated on a quantum computer. However, as the operation $\Sigma$ is a diagonal matrix with non-negative entries (also called singular values), this matrix is not necessarily unitary and simulation of such an evolution is not possible on a quantum computer using unitary gates. Moreover, the singular values need not always be less than one. The requirement of singular values to be less than or equal to 1 arises from the fact that they are used to parameterize unitary gate in the required circuit. Consider $a$ to be a singular value of $\Sigma$. Then the corresponding parameterized unitary gate can be written in the following form, iff $a \leq 1$:
\begin{equation}
U(a)=\begin{pmatrix}
a & -\sqrt{1-a^2}\\
\sqrt{1-a^2} & a
\end{pmatrix}\;.
\end{equation}

(b) {\it $L^2$-normalization:} We will next $L^2$-normalize the $NUTE$ matrix using the maximum singular value, $\max(\Sigma)$. We denote this normalized form as $NUTE_N$. The singular values of a such a $L^2$-normalized matrix (of 2 dimensions) are always $1$ and $m$. where $m \leq 1$ (equality arising when $NUTE$ is unitary). Note that such a normalization does not affect the output state as the output is not affected when the transformation is multiplied by a scalar constant.

(c) {\it Introduction of an ancilla:}  Simulating  the $NUTE_N$ in its current form still may not be possible in a quantum computer. To circumvent this issue, we next dilate the Hilbert space of the system by using an ancillary system. Note that the joint operation of the system and ancilla can be unitary, while that for the system itself remains non-unitary. We start with the joint state $|\xi\rangle|0\rangle$ of the system and the ancilla.

(d) {\it Two-qubit operation:} We create a two-qubit gate $C\Sigma = U(1) \oplus U(m) $, where $U(1)=I$ ($I$ is an identity matrix) and $U(m)$ is also a unitary matrix. Note that  $C\Sigma$ is unitary because the direct sum of unitary matrices is always unitary.

Now we consider the operation
\begin{equation}
(U \otimes I)[C\Sigma](V^{\dagger} \otimes I)
\end{equation} 
acting on the state $|\xi\rangle|0\rangle$. This is described in Fig. \ref{circuit}(a). This operation can be written as $(U \otimes I)[(\Sigma \otimes I)+(\Sigma^{'} \otimes ZX)](V^{\dagger} \otimes I)$, because,
\begin{widetext}
\begin{equation}
C\Sigma=U(1) \oplus U(m)
= \begin{pmatrix}
1 & 0 & 0 & 0 \\
0 & 1 & 0 & 0\\
0 & 0 & m & -\sqrt{1-m^2}\\
0 & 0 & \sqrt{1-m^2} & m \\
\end{pmatrix}
=(\Sigma \otimes I)+(\Sigma^{'} \otimes ZX)\;
\end{equation}
\end{widetext}
where
\begin{equation}
    \Sigma^{'}=\begin{pmatrix}
0 & 0\\
0 & \sqrt{1-m^2}\\
\end{pmatrix}\;.
\end{equation}

Here $Z$ and $X$ are usual Pauli spin operators. 

In presence of the ancilla, the operator $NUTE_N$ can then be revised as $U(C\Sigma)V^\dag$. The output of this operation on $|\xi\rangle|0\rangle$ is then given by 
\begin{equation}(U\Sigma V^{\dagger}|\xi\rangle)|0\rangle - (U \Sigma^{'}V^\dag|\xi\rangle)|1\rangle=|\phi,0\rangle+|\phi^{'},1\rangle\;,
\label{opstate}
\end{equation}
where $|\phi^{'}\rangle=-U\Sigma^{'}V^\dag|\xi\rangle$. 

(e) {\it Post-selection:} We apply a post-selection protocol on the above output state such that the ancilla is projected into $|0\rangle$. This can be represented by the following expression: 
\begin{equation}(I \otimes P_0)(|\phi,0\rangle+|\phi^{'},1\rangle) = |\phi,0\rangle\;,
\end{equation}
where $P_0 = |0\rangle\langle0|$. Upon normalization (which is performed by the measurement operation itself), we have the state $\frac{1}{\mathcal N}|\phi,0\rangle$ which is the required output from this non-hermitian evolution. Here ${\mathcal N}$ is  given by \begin{equation}
    {\mathcal N}=\sqrt{(\langle\phi,0|+\langle\phi^{'},1|)(I\otimes P_0)(|\phi,0\rangle+|\phi^{'},1\rangle)}=\sqrt{\langle\phi,0|\phi,0\rangle}\;.
\end{equation}
The probability of success of the post-selection protocol  can be calculated as  
\begin{equation}
\left|\frac{\langle\phi|\phi\rangle}{\sqrt{|\langle\phi|\phi\rangle|^2+|\langle\phi^{'}|\phi^{'}\rangle|^2}}\right|^2\;.
\end{equation}
We apply the repeat-until-success strategy to extract out the $|\phi^{'},1\rangle$ state through post-selection.

(f) {\it Calculation of the QFI:} We intend to find the $QFI$ for parameter estimation using a two-level non-hermitian sensor.  We choose one of the eigenstates, denoted by $|\phi_{+}(\gamma) \rangle$, of the two-level system. To calculate $QFI$,  we first obtain the expression of the fidelity as
\begin{equation}
    F=|\langle\tilde{\phi}_+(\gamma+\delta \gamma)|\phi_+(\gamma)\rangle|^2 \;,
    \label{Fidd}
\end{equation}
where $\langle\tilde{\phi}_+(\gamma+\delta \gamma)|=\langle\chi_+(\gamma + \delta \gamma)|$ is the left eigenvector, corresponding to the right eigenvector $|\phi_+(\gamma + \delta \gamma)\rangle$, according to the bi-orthogonality rules [see Appendix B]. The fidelity can be found on the actual quantum computing hardware using a SWAP test, as described in Fig. \ref{circuit}(b) \cite{buhrman2001quantum}.  The $QFI$ is then calculated numerically by using the equation 
\begin{equation}
    QFI =  \lim_{\delta \gamma \rightarrow 0}\frac{D_b^2}{4(\delta \gamma)^2}\;,
\end{equation} 
where $D_b^2=2-2F$ is known as Bures distance \cite{braunstein1994statistical,paris2009quantum}.

\section{\label{sec:3} SIMULATING EP SENSING DYNAMICS ON QC}

In this Section, we show how to implement the different steps to simulate a non-unitary evolution based on a non-Hermitian Hamiltonian, as mentioned in the previous Section, using the IBM Q Experience. We will investigate two different Hamiltonians for EP sensing. We start with the following Hamiltonian:
\begin{equation}
NH_1=\begin{pmatrix}
0 & 0.5+\gamma\\
0.5-\gamma & 0 
\end{pmatrix}\;,
\label{Ham}
\end{equation}
where $\gamma \nless 0$ is the parameter that is to be estimated in this protocol. Therefore, the protocol in Sec. II becomes more relevant when the non-unitary operation, that the state of the system is subjected to, is unknown, or in other words, the single-parameter Hamiltonian is rather unknown.

The matrix (Eq. \ref{Ham}) is non-Hermitian as $NH_1 \neq (NH_1)^{\dagger}$. The eigenvectors of $NH_1$ are (after using normalization condition for bi-orthogonal systems: $\langle\phi|\chi\rangle=1$) $|\phi_{\pm}\rangle=\frac{1}{\sqrt{2}} \left[\pm\frac{\sqrt{1-4\gamma^2}}{2\gamma-1}, 1\right]^T$ and those of $(NH_1)^{\dagger}$ are $|\chi_{\pm}\rangle=\frac{1}{\sqrt{2}}\left[\mp\frac{\sqrt{1-4\gamma^2}}{2\gamma+1}, 1\right]^T$, where $T$ represents transposition. It should also be noted that $\langle\phi_{\pm}|\phi_{\mp}\rangle \neq 0$ unlike in the case of hermitian operators (the same applies for eigenvectors of $(NH_1)^{\dagger}$, as well). However $\langle\phi_{\pm}|\chi_{\mp}\rangle = 0$. This is further elaborated in the Appendix B. We observe that this system exhibits the exceptional point degeneracy at $\gamma=0.5$. Note that the eigenvalues of $NH_1$ are given by $\pm \sqrt{0.25-\gamma^2}$.

The $QFI$ for a single parameter $\gamma$ in a bi-orthogonal system is defined as 
\begin{eqnarray}
QFI_{\gamma}=4(\langle\partial_{\gamma}\tilde{\phi}_{+}|\partial_{\gamma}\phi_{+}\rangle-|\langle\partial_{\gamma}\tilde{\phi}_{+}|\phi_{+}\rangle|^2)\\
=4(\langle\partial_{\gamma}\chi_{+}|\partial_{\gamma}\phi_{+}\rangle-|\langle\partial_{\gamma}\chi_{+}|\phi_{+}\rangle|^2)\; \label{qfi1}
\end{eqnarray}
where $|\partial_{\gamma}\phi\rangle=\frac{\partial|\phi(\gamma)\rangle}{\partial \gamma}$ and $| \partial_{\gamma}\chi\rangle=\frac{\partial|\chi(\gamma)\rangle}{\partial \gamma}$. Note that this equation matches the $QFI$ equation used in step (f) of Sec. II.

It can be calculated using (Eq. \ref{qfi1}) that for the Hamiltonian Eq. \ref{Ham}, 
\begin{equation}
QFI_{\gamma}=4/(4\gamma^2-1)^2\;.
\end{equation}
We can easily see that the $QFI_\gamma$ diverges at $\gamma=0.5$, which is the same value of the parameter $\gamma$, at which the EP is attained in the system, Fig. \ref{result1}. As mentioned in the Introduction, this divergence is achieved generally at EPs \cite{kato2013perturbation,brody2013information}, despite some converse result  \cite{chen2019sensitivity}.  We emphasize that at the EP degeneracies, the perturbed eigenvalues and eigenvectors must be  expanded in terms of Newton-Puiseux series \cite{kato2013perturbation} with fractional exponents of the perturbation strengths rather than Taylor series, to get the required divergence [see Appendix C]. 

We have performed simulation of the above system using ``state\_vector simulator" and "qasm\_simulator" \cite{state2019} and compared its  results with the theoretical result of the $QFI_\gamma$ of the system. The statevector\_simulator emulates an ideal operation of a quantum circuit and provides the quantum state vector of the system after simulation concludes rather than measurement outcomes. Ideally a quantum circuit is not capable of providing information of a state directly.  However, such a simulator has valuable applications in theoretical analysis and sometimes troubleshooting of algorithms and hence is ideal for such situations. Our simulation can also be completely performed using only measurement counts (unlike the state\_vector simulation) and the divergence can still be obtained as shown in Fig. \ref{result1} by using the "qasm\_simulator". The qasm\_simulator is devised for simulating how a quantum computer would exactly output it's result. This shows that the system can also be easily replicated on an actual hardware. The Fig. \ref{result1} shows how $\log(QFI_\gamma)$ varies with $\gamma$. It can be seen that the $QFI_\gamma$ indeed diverges at the EP $\gamma=0.5$.

We next simulate the second Hamiltonian as in \cite{chen2019sensitivity}:
\begin{equation}
NH_2=\begin{pmatrix}
\frac{1}{2}(\gamma-i\epsilon) & g\\
g & -\frac{1}{2}(\gamma-i\epsilon) 
\end{pmatrix}\;.
\label{Ham2}
\end{equation}
There exists an EP at $\gamma = 0$, when $g=|\epsilon|/2$. It can be seen that the $QFI_\gamma$ diverges at $\gamma=0$. The plot of $\log(QFI_\gamma)$ with respect to $\gamma$ is given in Fig. \ref{result2}. If we consider $\epsilon=2$ and $g=1$ and use the perturbation methods mentioned in \cite{brody2013information} we find that $QFI_{\gamma} \propto \frac{1}{4\gamma^2}$. It diverges at $\gamma=0$ which leads us to an EP of $NH_2$. 

The results for both the Hamiltonians (Eq. \ref{Ham}) and (Eq. \ref{Ham2}) demonstrate the claim that the $QFI_\gamma$ diverges at EPs.

\subsection{Multiple EP}
There also exist certain systems that exhibit multiple EPs \cite{Ding-2016, Lee-2012, Pap-2018}.  Let us simulate a two-level system with multiple exceptional points, described by the following Hamiltonian, as an example:
\begin{equation}
NH_3=\begin{pmatrix}
1 & \sin(10\gamma\pi) \\
\cos(10\gamma\pi) & 1
\end{pmatrix}
\label{Ham3}
\end{equation}

The eigenvalue of this system are $\lambda_{+,-}=\frac{1}{2}(2 \pm \sqrt{2 \sin(20\pi\gamma)})$. They correspond to the normalized eigenvectors of $NH_3$, given by $|\phi_{+,-}\rangle=\frac{1}{\sqrt{2}}
\left[
\pm\frac{\sqrt{2} \sin(10\pi\gamma)}{\sqrt{\sin(20\pi\gamma)}},1\right]^T$. Note that The eigenvalues of $NH_3^{\dagger}$ are complex conjugates to those of $NH_3$, with the corresponding normalized eigenvectors
$|\chi_{+,-}\rangle=\frac{1}{\sqrt{2}}
\left[
\pm\frac{\sqrt{2} \cos(10\pi\gamma)}{\sqrt{\sin(20\pi\gamma)}},1
\right]^T$. The QFI is then calculated as
\begin{equation}
QFI_{\gamma}=\frac{200\pi^2}{\sin^2(20\pi\gamma)}\;.
\end{equation}

\begin{figure*}
    \centering
    \includegraphics[scale=0.8]{ 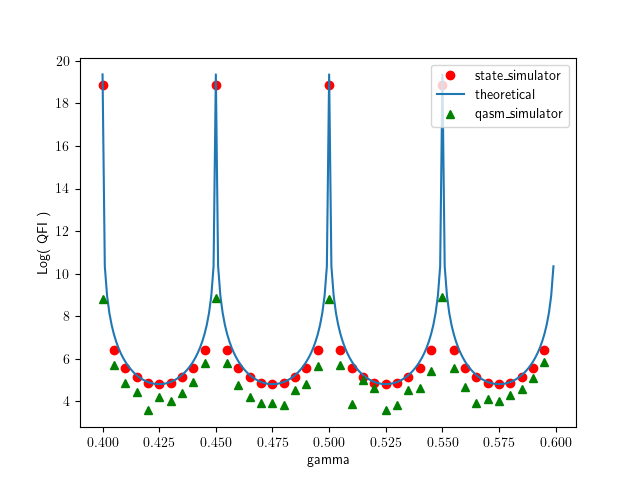}
    \caption{The variation of $\log(QFI)$ with respect to  $\gamma$ for the Hamiltonian $NH_3$ (Eq. \ref{Ham3}). The $QFI$ diverges at the EP, which appears at $\gamma=0.05a$, where we have chosen the integer $a\in [8,12]$. Blue curve represents theoretical plot and red dots indicate simulation results for various values of $\gamma$.}
    \label{result3}
\end{figure*}

The system is simulated and the QFI is plotted in Fig. \ref{result3}. It can seen that there are multiple EP points at $\gamma=a \times  0.05$ where $a \in I$. 

\section{\label{sec:4} VARIOUS NOISE MODELS ON EP SENSORS}

Even though the $QFI$ diverges at EP, it still remains to be explored how noise affects the $QFI$. Let us consider various noise models and calculate the maximum $QFI$ that can be attained for various parameters controlling the noise. In this section we will use the 'state\_vector simulator' for simulation.

\begin{figure*}
     \centering
     \begin{subfigure}
         \centering
         \includegraphics[scale=0.6]{ 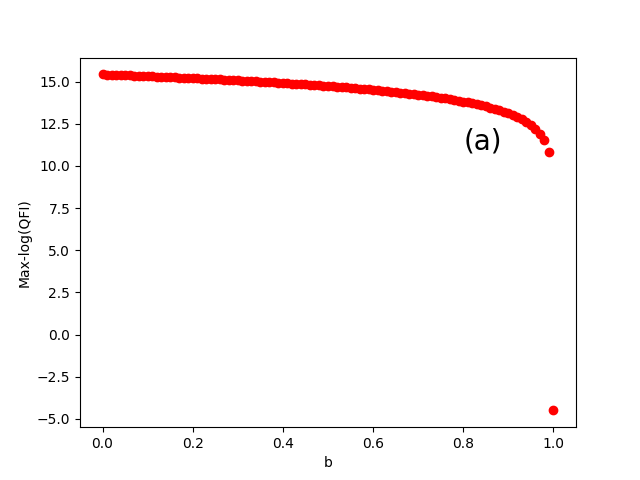}
     \end{subfigure}
     \hfill
     \begin{subfigure}
         \centering
         \includegraphics[scale=0.6]{ 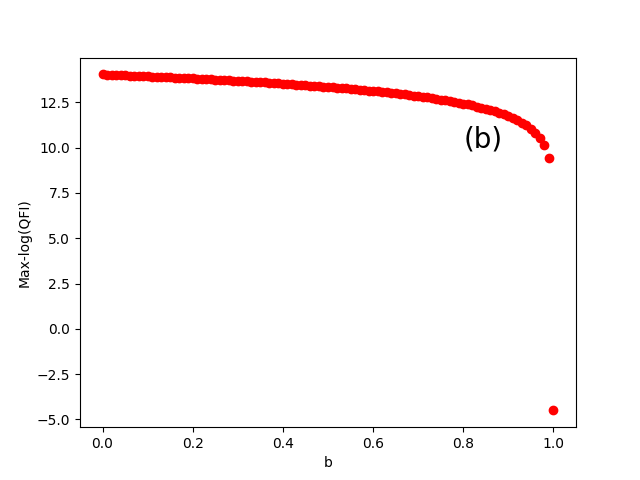}
     \end{subfigure}
    \hfill
        \caption{The variation of $\max[\log(QFI)]$ at EP with respect to the amplitude damping parameter $b$ (a) for $NH_1$ and (b) for $NH_2$}
        \label{ADfig}
\end{figure*}

Considering the amplitude damping noise that is parameterized with a decay probability $b$, a noiseless density matrix $\rho$ transforms into the following matrix: 
\begin{equation}
    N_{AD}(\rho)=\begin{pmatrix}
    \rho_{00}+b\rho_{11} & \sqrt{1-b}\rho_{01}\\
    \sqrt{1-b}\rho_{10} & (1-b)\rho_{11}
    \end{pmatrix}\;.
    \label{AD}
\end{equation}
We show the variation of $\max[\log(QFI)]$ (i.e., the maximum of natural logarithm of $QFI$) with respect to the decay rate $b$ at the EP $\gamma=0.5$ for $NH_1$ [Fig. \ref{ADfig} (a)] and $\gamma = 0$ for $NH_2$  [Fig. \ref{ADfig} (b)].  It can be observed that the $\max[\log(QFI)]$ remains significantly high for a wide range of $b$: $0\le b<1$, except at $b=1$, where the function suddenly drops to a negative value, i.e., where $QFI$ becomes less than unity. Thus, $b=1$ leads to a large upper bound of the error of parameter estimation [see Eq. (\ref{delgam})]. 

Next we consider the Pauli noise, which can be characterized by three types of errors, namely, the bit-flip, bit-phase-flip, and phase-flip errors, represented by the Pauli matrices X, Y, and Z, respectively. In presence of Pauli error, a noiseless density matrix $\rho$ 
 transforms into the following form:  
\begin{equation}
    N_{Pauli}(\rho)=(1-p)\rho+p_1X\rho X+p_2Y\rho Y +p_3Z\rho Z\;,
    \label{Pauli}
\end{equation}
where $p$ is the probability of no error, $p_i$ $(i=1,2,3)$ is that of X, Y, and Z error, respectively, and the condition $\sum_{i=1}^3p_i=p$ preserves the normalization of the density matrix. 

\begin{figure*}
     \centering
     \begin{subfigure}
         \centering
         \includegraphics[scale=0.5]{ 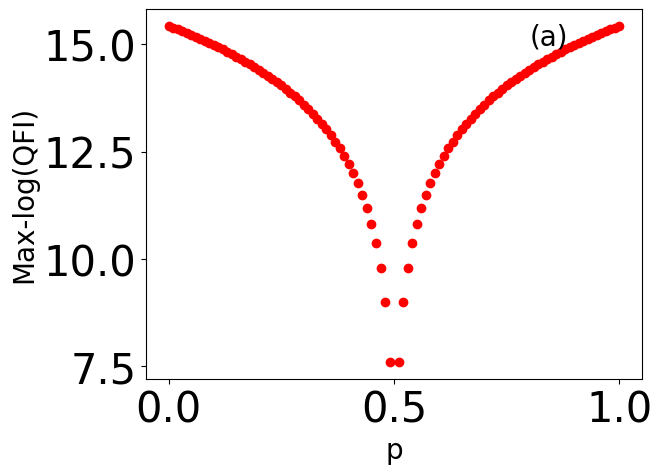}
         \label{fig:y equals x1}
     \end{subfigure}
     \hfill
     \begin{subfigure}
         \centering
         \includegraphics[scale=0.5]{ 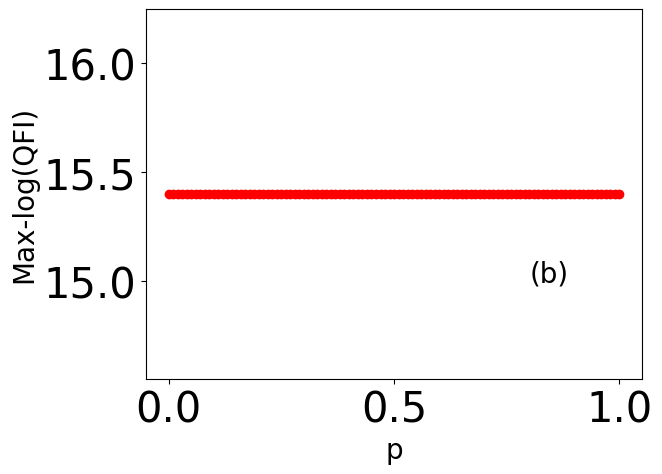}
         \label{fig:three sin x1}
     \end{subfigure}
     \hfill
     \begin{subfigure}
         \centering
         \includegraphics[scale=0.5]{ 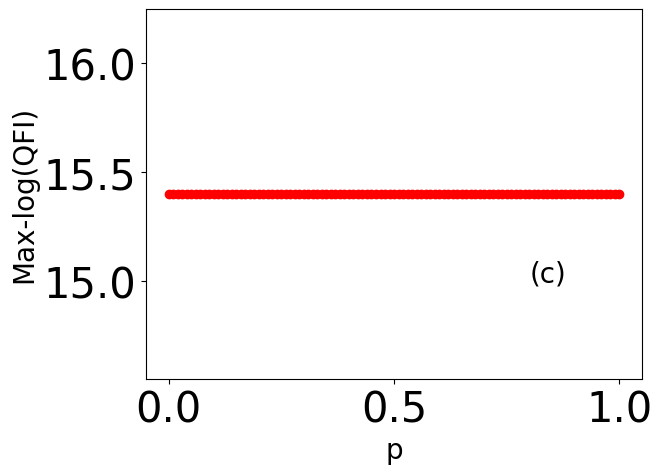}
         \label{fig:five over x1}
     \end{subfigure}
     \hfill
     \begin{subfigure}
         \centering
         \includegraphics[scale=0.5]{ 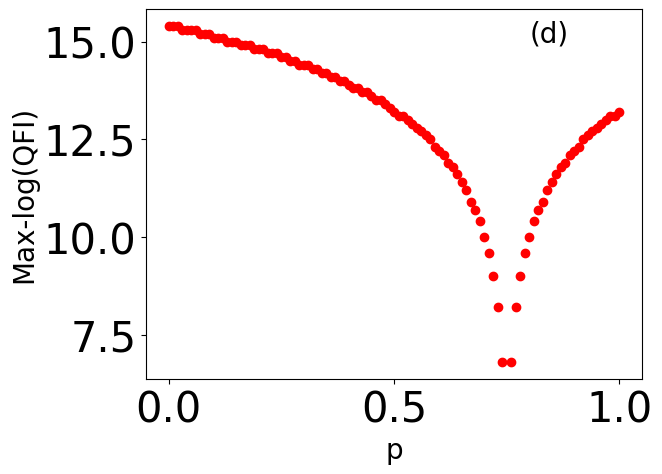}
         \label{fig:five over x2}
     \end{subfigure}
     \hfill
        \caption{The variation of $\max[\log(QFI)]$ with respect to $p$ for different Pauli errors for $NH_1$. (a) Bit-Flip error: $p_1=p$, $p_2=0$, $p_3=0$. (b) Bit-Phase-Flip error: $p_1=0$, $p_2=p$, $p_3=0$. (c) Phase flip error: $p_1=0$, $p_2=0$, $p_3=p$. (d) All errors with equal probability: $p_1=p_2=p_3=p/3$.}
        \label{Paulierror1}
\end{figure*}

We show in Fig. \ref{Paulierror1} how the $\max[\log(QFI)]$ varies with $p$, at the EP $\gamma=0.5$ for $NH_1$. We can clearly see from Fig. \ref{Paulierror1}(a) that the bit-flip error affects the maximum achievable $QFI$ to a large extent. The minimum value reached at $p=0.5$ is 0. On the other hand, the Y and Z errors do not effect the $QFI$ at all, as can be seen in Figs. \ref{Paulierror1}(b) and (c). In the case of the equal probability $p/3$ of all three errors, however, the value of $p$ is asymmetrically located at 0.75 in the range [0,1], where the maximum value of the $QFI$ becomes  unity and its logarithm vanishes. For other values of $p$, in this case, the $QFI$ can reach a much larger value leading to a far better EP-assisted sensing. 

\begin{figure*}
     \centering
     \begin{subfigure}
         \centering
         \includegraphics[scale=0.5]{ 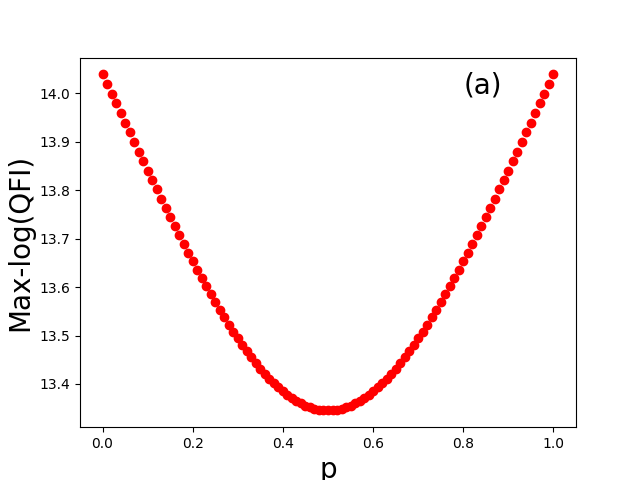}
     \end{subfigure}
     \hfill
     \begin{subfigure}
         \centering
         \includegraphics[scale=0.5]{ 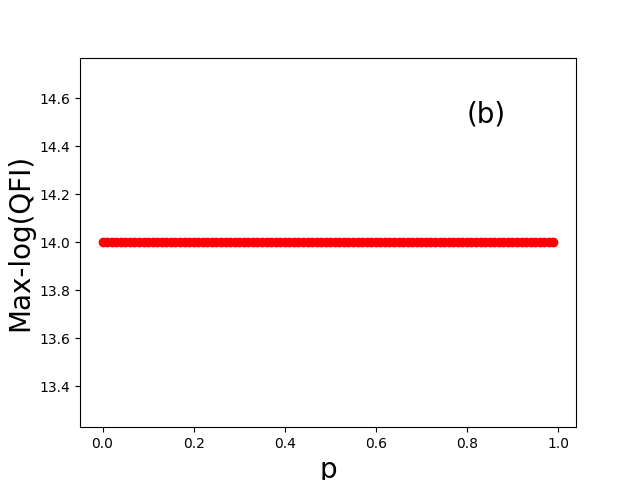}
     \end{subfigure}
     \hfill
     \begin{subfigure}
         \centering
         \includegraphics[scale=0.5]{ 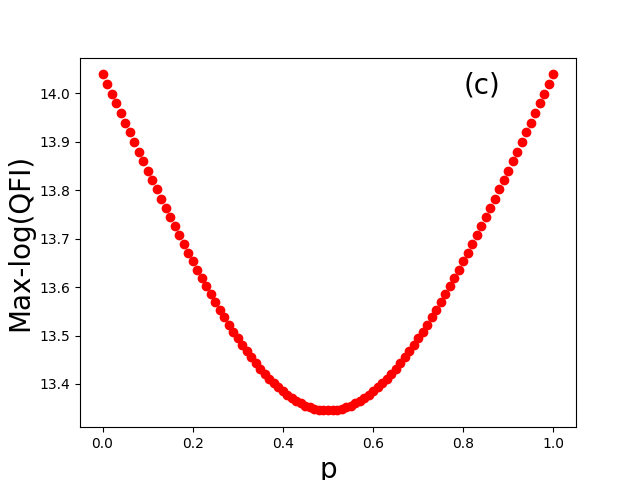}
     \end{subfigure}
     \hfill
     \begin{subfigure}
         \centering
         \includegraphics[scale=0.5]{ 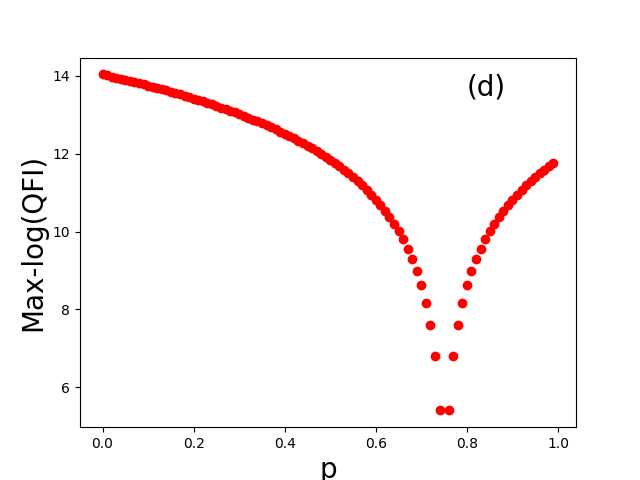}
     \end{subfigure}
     \hfill
        \caption{The variation of $\max[\log(QFI)]$ with respect to $p$ for different Pauli errors for $NH_2$. (a) Bit-Flip error: $p_1=p$, $p_2=0$, $p_3=0$. (b) Bit-Phase-Flip error: $p_1=0$, $p_2=p$, $p_3=0$. (c) Phase flip error: $p_1=0$, $p_2=0$, $p_3=p$. (d) All errors with equal probability: $p_1=p_2=p_3=p/3$.}
        \label{Paulierror2}
\end{figure*}

In Fig. \ref{Paulierror2}, we show the variation of $\max[\log(QFI)]$ with $p$, at the EP $\gamma=0$ for $NH_2$. It is obvious from the Figs. \ref{Paulierror2}(a) and (c) that the bit-flip  and phase flip errors have a minimal effect on the maximum achievable $QFI$. Interestingly, unlike the case for $NH_1$, the $\max[\log(QFI)]$  does not vanish for any $p$. In both these cases, the value of parameter $p$ at which it reaches minima is 0.5, but in the case of $NH_2$, it remains as high as 13.35.  On the other hand, the Y error does not effect the $QFI$ at all, as can be seen in Fig. \ref{Paulierror2}(b). Therefore, our protocol stays robust against these errors. When all the three errors affect the system equally probably, the dependence of $\max[{\log (QFI)}]$ on $p$ exhibits the similar behavior as in the case of $NH_1$ [see Fig. \ref{Paulierror2}(d)].

Comparing Fig. \ref{Paulierror1} and \ref{Paulierror2} it can be deduced that in presence of noise with a parameter $a$, we need to choose the non-hermitian system $NH$ such that the $QFI_{\gamma}(a_0)) \gg 0$, where $a_0$ is the value of the noise parameter at which the $QFI$ reaches minima, and $\gamma$ is the parameter to be estimated.  

\section{\label{sec:5} Physical example of quantum sensing using EPs}

Consider that there is a vector signal to be estimated such that it is given as $\overrightarrow{V}=(V_x,V_y,V_z)$. Consider $V_y=V_z=0$ and the x component of the signal is non-zero, for simplicity of the demonstration. Hence, this will be an example of Rabi sensing using EPs, as $V_x$ will appear in the cross terms of the total hamiltonian. Sensing of such vector signals using hermitian sensing quantum systems has been discussed in \cite{Degen-2017}. 

In a general quantum sensing protocol, the effect of this signal is acted upon as a perturbation to the dynamics of a sensing system. Through this the perturbations are imprinted on the population dynamics of the sensing system and a measurement of the populations will reveal the properties of the signal by comparing them to theoretical estimates of the population upon perturbation. The EP quantum sensor will perform in the same way, except with a few changes.

The hamiltonian of the signal discussed above will be 

\begin{equation}
H_V=\frac{\epsilon}{2}V_x\sigma_x
\label{signalham}
\end{equation}

Here $\epsilon$ is the transduction constant and $\sigma_x$ is the Pauli-X matrix. 

Consider a physical system such as \cite{naghiloo2019quantum}. We reproduce their physical system in this article and use it to showcase the feasibilty of using this system as EP sensor for Rabi sensing with very high precision. It considers a three level system created using transmon circuit embedded in a cavity. Let the states of the system be given by $|g\rangle,|e\rangle,|f\rangle$ in respective order of their energies. Consider that the population loss rate from $|f\rangle$ to $|e\rangle$ is $\gamma_f$ and from $|e\rangle$ to $|g\rangle$ as $\gamma_e$. The value of $\gamma_e>>\gamma_f$. This is achieved in the article by using impedance mismatch element. In such a case the hamiltonian of the sub-system of first two excited states ($|f\rangle$ and $|e\rangle$) is given as 

\begin{equation}
H_0=\tilde{J}\sigma_x+\Delta I-i\frac{\gamma_e}{2}|e\rangle\langle e|
\end{equation}

We can see that using such a system we can work entirely in the sub-system of the three level system. Here $\tilde{J}$ is the coupling constant between the first two excited states and $\Delta$ the detuning parameter. After perturbing the above system by the signal hamiltonian described in Eq. \ref{signalham}, we get the total hamiltonian of the system to be 

\begin{equation}
H=H_0+H_V=J\sigma_x+\Delta I -i\frac{\gamma_e}{2}|e\rangle\langle e|
\label{entiresys}
\end{equation}

where $J=\tilde{J}+\frac{\epsilon}{2}V_x$. 

The eigenvalues of the system are 

\begin{equation}
\lambda_{+,-}=\Delta-\frac{i\gamma_e}{4}\pm\frac{d}{4}
\end{equation}

and the eigenvectors are 

\begin{equation}
|\phi_{+,-}\rangle=\frac{1}{\sqrt{2}}
\begin{pmatrix}
(-i\gamma_e\pm d)/4J\\
1
\end{pmatrix}
\end{equation}

here $d=4\sqrt{J^2-(\frac{\gamma_e}{4})^2}$. The state of the system over time is then given as 

\begin{equation}
|\phi_e(t)\rangle=\frac{-2 J}{d}\left(e^{-i \lambda_- t}\left|\phi_-\right\rangle-e^{-i \lambda_+ t}\left|\phi_+\right\rangle\right)
\label{phie}
\end{equation}

\begin{equation}
\left|\phi_f(t)\right\rangle=\frac{ \sqrt{128} J^2}{d(d +i \gamma_e)}\left(e^{-i \lambda_- t}\left|\phi_-\right\rangle+C e^{-i \lambda_+ t}\left|\phi_+\right\rangle\right)
\label{phif}
\end{equation}

Here $C=\frac{d(d+i\gamma_e)-8 J^2}{8 J^2}$
It should be noted that sum of the populations ( $P_e=|\langle \phi_e(t)| \phi_e(t) \rangle|^2$, $P_f=|\langle \phi_f(t)| \phi_f(t) \rangle|^2$) in the sub-system is not 
a constant due to the interaction with the ground state $|g\rangle$. 

The populations can then be normalized by dividing by $P_e+P_f$ such that the normalized populations are 

\begin{equation}
P_f^{\mathrm{n}}=\frac{P_f}{P_f+P_e}
\label{pope}
\end{equation}
\begin{equation}
P_e^{\mathrm{n}}=\frac{P_e}{P_f+P_e}
\label{popf}
\end{equation}

It can seen from Eq. \ref{phie} to \ref{popf} that they are dependent on $\Delta,\gamma_e,J$. 

Now performing a measurement of either of the population $P_f$ (or $P_e$) over time and normalizing it using post-selection \cite{naghiloo2019quantum}, we can compare the experimentally obtained population with the theoretical expressions and using the known parameters of the sensing system, we can deduce the unknown parameter (in this case $\epsilon V_x$. In this particular case the known parameters are $\tilde{J}$, $\Delta$ and $\gamma_e$.

The post-selection in the experiment can be performed by measuring the ground state population $P_g$ over time and then using it with the conservation of population formula i.e. $P_e+P_f=P-P_g$. 

The entire procedure mentioned above resembles the procedure for Rabi sensing, except 1) a non-hermitian sub-system is used instead of a hermitian system, 2) post-selection is performed over time to normalize the populations of the sub-system over time. 

Now, let us understand how this is advantageous i.e. what happens to such a sensing mechanism at EP. The entire system (Eq. \ref{entiresys}) is at EP when $\Delta=0$ and when $\epsilon V_x= \frac{\gamma_e}{2}-2\tilde{J}=L$. The values of $\gamma_e$ and $\tilde{J}$ can be tuned as performed in \cite{naghiloo2019quantum}).
Now, if for a particular setting of $\gamma_e$ and $\tilde{J}$, the above condition is met, then $\Delta V_x\rightarrow 0$, due to divergence of $QFI_{V_x}$ at EP for the subsystem and the QCRB (Eq. \ref{delgam}). This fact has been effectively demonstrated by simulations using the circuit model in previous sections and has also been derived for a general hamiltonian at EP in Appendix. \ref{AppC}.  In fact the divergence of the $QFI$ has also been observed in the work by \cite{naghiloo2019quantum} where the QFI is plotted w.r.t $\tilde{J}$ instead of $V_x$. 

We should be aware that such a divergence only occurs at a particular point (i.e. iff $\epsilon V_x=L$ for our case). This is because we consider a system with isolated EPs. However, if we use a system with Exceptional surface (ES) instead of EP, we could get such a high precision for a range of values of $V_x$. In fact such proposals are already available in optical systems \cite{Zhong-2019,qin2021experimental,zhang2019experimental}. 

Similarly, Ramsey sensing (measuring $V_z$ from the vector signal) can also be performed by using the above system. In this case the value of $V_z$ will affect the detuning parameter $\Delta$ instead of the cross-terms $J$. Now operating the system at EP and using the similar methods above for Rabi sensing at EP, we can estimate the value of $V_z$ with high precision.

\section{\label{sec:6} Limitations of ancilla-based method for quantum sensing simulation}

Let us now discuss the limitations in the ancilla-based method for EP sensing simulation.

First, we should mention the necessity to normalize the non-unitary operator to implement it on a quantum computer. As discussed in step (b) in Sec. II, the normalized non-unitary time evolution operator $NUTE_N$ is related to the original time evolution operator $NUTE$ as:
\begin{align}
    NUTE_N=\frac{1}{\max(\Sigma)}NUTE\;,
\end{align}
where $\max(\Sigma)$ is the maximum of the singular value of the matrix. This means the actual transformation is:
\begin{align}
    NUTE_N|\xi\rangle=|\phi^{''}\rangle=\frac{1}{\max(\Sigma)}|\phi\rangle\xrightarrow[]{Norm}\frac{1}{\mathcal N}|\phi\rangle,
\end{align}
where $\frac{1}{\mathcal N}|\phi\rangle$ is the required state from the original non-unitary time evolution.

Second, the post-selection puts a resource overhead on the simulation of the system being studied. The probability of success of finding the ancilla  to be in $|0\rangle$ state is uncontrollable, as it can be seen from the previous Section. For every unsuccessful attempt, the depth of the circuit is increased by a factor of 3. There are various methods to reduce the resource overhead, e.g., ancilla thermalization \cite{wright2021automatic}. There are some proposals where non-unitary evolutions can be performed using dissipation engineering and without post-selection \cite{zapusek, nirala2019measuring, wang2019non}. A study of link between weak measurements and sensing at EPs of the non-hermitian systems is promising \cite{aharonov1988result,tanaka2013information, gray2009post,abbott2019anomalous}. 

Third, note that these methods only apply to second-order exceptional point systems. For higher-dimensional systems including qutrits and qudits, the circuits need to be modified. Higher order EPs have been shown to have higher sensitivity and being more robust \cite{hodaei2017enhanced}.

\section{\label{sec:7} Conclusion}
We demonstrate in this paper how a unitary circuit can be used to study non-hermitian sensors, thus solving the need of developing new hardware or experiments to study these systems. We proposed use of ancillary system in this regard.  We also showcase the divergence of $QFI$ at exceptional points using three different Hamiltonians. We study the effect of noise on one such system and show that the $QFI$ at the EP stays robust against all the Pauli and amplitude errors for a large range of parameters. For certain errors, namely X error, the $QFI$ can be ameliorated using quantum error correction techniques. We also direct to the work of \cite{naghiloo2019quantum} as an example of physical system for quantum sensing at EP. 

\begin{acknowledgments}
One of us (C.W.) would like to acknowledge Council of Scientific and Industrial Research (CSIR), India for financial assistance through CSIR-JRF fellowship. 
\end{acknowledgments}

\section*{Author Declarations}
\subsection*{Conflict of interest}
The authors have no conflicts to disclose.

\section*{Data Availability Statement}
The data that support the findings of this study are available from the corresponding author upon reasonable request.

\appendix

\section{Puiseux Series}
Puiseux series,  are used to represent functions that have algebraic singularities, which can include branch points and other types of singularities. Puiseux series have terms that include fractional powers of the variable, as well as integer powers and constant terms.  The coefficients in a Puiseux series may depend on the path taken to approach the singularity, whereas the coefficients in a Laurent series depend only on the singularity itself.

Let us consider a two-level system with the Hamiltonian in the Jordan-Block (without loss of generality) form $H_{EP}$, that is  perturbed by a general Hamiltonian $\epsilon H_1$:
\begin{align}
    H=
    \begin{pmatrix}
    \lambda & 1\\
    0 & \lambda
    \end{pmatrix} + \epsilon\begin{pmatrix}
    x & y^{*}\\
    y & x+2z
    \end{pmatrix}=H_{EP}+\epsilon H_1\;.
\end{align}
Consider $\lambda^{'}_{EP}$ as eigenvalues of the perturbed Hamiltonian from EP. 
\begin{align}
    \lambda^{'}_{EP}=(\lambda+\epsilon(x+z)) \pm \sqrt{\epsilon^2(z^2-|y|^2)+\epsilon y^*}\;.\label{eigep}
\end{align}
The second term can be expanded in Puiseux series. This is due to the term $\epsilon y^*$. Such a case does not occur if the original Hamiltonian (in diagonal form, without loss of generality) is hermitian $H_H$. The corresponding eigenvalues would read as  
\begin{align}
    \lambda^{'}_{H}=(\lambda+\epsilon(x+z)) \pm \sqrt{\epsilon^2(z^2-|y|^2)}\;.
\end{align}

Below, we provide an example of the Puiseux series: Let us consider the function $f(x)=\sqrt{x^2-1}$. Using the change of variable, $y=x-1$, we have $f(y)=\sqrt{y^2+2y}$. Expanding about $y=0$, we have  $f(y)=\sqrt{2y}(1+\frac{y}{4}-\frac{y^2}{32}+...)$. Substituting the expression of $y$ back, we get
\begin{align}
    f(x)=\sqrt{2}\sqrt{x-1}+\frac{2}{4}(x-1)^{3/2}+\frac{\sqrt{2}}{32}(x-1)^{5/2}+...\;.
\end{align}
which is a Puiseux series expansion. 

Eigenvectors of a system perturbed from an exceptional point, as in (Eq. \ref{eigep}), follow a similar series expansion.

 Note there are subtle differences between Puiseux series and Laurent series. One needs to consider them while expanding the eigenvectors and eigenvalues of a Hamiltonian around EPs. Puiseux series are useful in studying the behavior of functions at singular points, e.g., the  algebraic functions, which have singularities at roots of their defining polynomials. Laurent series are used to represent functions that have poles, which are isolated singularities of the function where it becomes infinite. Laurent series have terms that include negative powers of the variable, as well as positive powers and constant terms. 
 The coefficients in a Puiseux series may depend on the path taken to approach the singularity, whereas the coefficients in a Laurent series depend only on the singularity itself.

 \section{Orthogonality, Bi-orthogonality, Self-orthogonality }

 Eigenvectors of non-hermitian (diagonalizable) systems do not follow orthogonality rules. Eigenvectors of non-hermitian non-diagonalizable matrices rather follow bi-orthogonality. Here, we will state bi-orthogonality rules without proof. 

Suppose $NH$ is the non-hermitian Hamiltonian and $(NH)^{\dagger}$ its hermitian conjugate. Suppose the eigenvalues of $NH$ are $\lambda_i$ ($i$ denoting the index of different eigenvectors of $NH$). The eigenvalues of $(NH)^{\dagger}$ are $\lambda_i^*$ (* denoting complex conjugate). If the eigenvectors of $NH$ are $|\phi_i\rangle$ and those of $NH^{\dagger}$ are $|\chi_i\rangle$, then the orthogonality rules for normalized eigenvectors are:
\begin{align}
\langle\chi_i|\phi_j\rangle=\delta_{ij}\;,\\
    \langle\chi_i|\chi_j\rangle \neq \langle\phi_i|\phi_j\rangle \neq \delta_{ij}\;.
\end{align}
This means that the left eigenvectors corresponding to $|\phi_i\rangle$ are not $\langle\phi_i|$, but $\langle\chi_i|$. Also, as in the case of corresponding eigenvalues,  $|\phi_i\rangle=(|\chi_i\rangle)^{*}$. 
The bi-orthogonality is a manifestation of time-irreversibility of non-unitary time evolutions as opposed to that of unitary time evolutions.

When the system Hamiltonian becomes non-diagonalizable, the bi-orthogonality rules also break down. The system becomes self-orthogonal. For simplicity, let us assume only exceptional points for two-level systems, such that there is possibility of only two-fold degeneracy. At the exceptional points, there is a degeneracy in eigenvectors too and hence there is also one eigenvector. Consider $\lambda_{EP}$ and $|\phi_{EP}\rangle$ to be eigenvalue and eigenvector of $NH_{EP}$. The eigenvalue of $NH_{EP}^{\dagger}$ is $\lambda_{EP}^{*}$ and the eigenvector $|\chi_{EP}\rangle$. As there is no second eigenvector existing for the eigenvalue, we consider what is known as generalized eigenvector. Such an eigenvector follows the equation as given below:
\begin{align}
H_{EP}|\phi_{EP}^{J}\rangle=\lambda|\phi_{EP}^J\rangle+|\phi_{EP}\rangle\\
H_{EP}^{\dagger}|\chi_{EP}^{J}\rangle=\lambda^{*}|\chi_{EP}^J\rangle+|\chi_{EP}\rangle\;.
\end{align}
where $J$ in superscript denotes generalized eigenvector. The self-orthogonality rules followed by such a system are as follows:
\begin{align}
    \langle\chi_{EP}|\phi_{EP}\rangle=\langle\chi_{EP}^J|\phi_{EP}^J\rangle=0\;.\\
    \langle\chi_{EP}|\phi_{EP}^J\rangle=\langle\chi_{EP}^J|\phi_{EP}\rangle=1\;.
\end{align}

\section{\label{AppC} Quantum Fisher Information at EPs}

The expressions for $QFI$ for non-hermitian diagonalizable Hamiltonians (irrespective of them having real eigenvalues) differ from those for the hermitian ones, at EP. We state the expressions for $QFI$ for hermitian system $H$ first.

Consider that the eigenfunction of $H(\gamma)$ is $\{|\psi_i(\gamma)\rangle\}$. Then, we have $\langle\psi_i(\gamma)|\psi_j(\gamma)\rangle=\delta_{ij}$. The $QFI$ can then be expressed as 
\begin{align}
    [QFI(\gamma)_{H}]_i=4(\langle\dot{\psi}_i(\gamma)|\dot{\psi}_i(\gamma)\rangle-|\langle\dot{\psi}_i(\gamma)|{\psi_i(\gamma)}\rangle|^2)\;.
\end{align}
where $|\dot{\psi}(\gamma)\rangle$ denotes $\frac{\partial|\psi(\gamma)\rangle}{\partial\gamma}$.
This equation then modifies to the following expression, for non-hermitian Hamiltonians $NH$,
\begin{align}
    [QFI(\gamma)_{NH}]_i=4(\langle\dot{\chi}_i(\gamma)|\dot{\phi}_i(\gamma)\rangle-|\langle\dot{\chi}_i(\gamma)|\phi_i(\gamma)\rangle|^2)\;,
\end{align}
where we have used the bi-orthogonality rules as expressed in Appendix B and the dot denotes partial differentiation with respect to the parameter as in hermitian case. 

Let us now consider a two-level system  at EP. The general form of $QFI$ for any generic system perturbed from EP has been derived in \cite{brody2013information,book:66568}. This perturbed Hamiltonian can be written as
\begin{equation}
    (NH)_{EP}+\gamma (NH)^{'}+....
\end{equation}
Here, $(NH)^{'}$ represents the perturbation Hamiltonian and $\gamma$ is the perturbation strength. Then according to the Puiseux series expansion \cite{kato2013perturbation} of the perturbed eigenvalues ($\lambda_p$) and eigenvectors ($|\phi_{p} \rangle$) around the algebraic singularity, i.e. the EP, we have:
\begin{equation}
    \lambda_{p}=\lambda+\gamma^{1/2}\lambda^{'}+\gamma \lambda^{''}+...
\end{equation}
and 
\begin{equation}
    |\phi_p \rangle= |\phi \rangle + \gamma^{1/2}|\phi^{'} \rangle+ \gamma |\phi^{''} \rangle....
\end{equation}
Using these expanded form, it can be calculated that the $QFI$ in general is given by \cite{brody2013information,book:66568} 
\begin{equation}
    QFI_{\gamma} \approx \frac{1}{4\gamma^2}\;,
\end{equation}
which shows divergence at $\gamma=0$ i.e. when there is no perturbation and the system is at EP for a two-level system with a second order exceptional point degeneracy.

\nocite{*}
\bibliography{aipsamp}

\end{document}